\documentclass[onecolumn,aps,prc,superscriptaddress]{revtex4}
\usepackage{amssymb}
\usepackage{amsmath}
\usepackage{graphicx}
\usepackage[normalem]{ulem}
\usepackage{multirow}
\usepackage{appendix}
\usepackage[usenames]{color}
\usepackage{bm}
\usepackage{hyperref}
\usepackage[utf8]{inputenc}

\begin{document}
\title{Uncertainty evaluation and correlation analysis of single-particle energies in phenomenological nuclear mean field: An investigation of propagating uncertainties for independent model parameters}

\author{Zhen-Zhen Zhang}
\affiliation{School of Physics and Microelectronics, Zhengzhou University, Zhengzhou 450001, China}

\author{Hua-Lei Wang}
\email[Corresponding author, ]{wanghualei@zzu.edu.cn}
\affiliation{School of Physics and Microelectronics, Zhengzhou University, Zhengzhou 450001, China}

\author{Hai-Yan Meng}
\affiliation{School of Physics and Microelectronics, Zhengzhou University, Zhengzhou 450001, China}

\author{Min-Liang Liu}
\affiliation{Institute of Modern Physics, Chinese Academy of Sciences,   Lanzhou
                                                                   730000, China
}

%\date{\today}
%============================================================================================================================================%
\begin{abstract}

 Based on Monte Carlo approach and conventional error analysis theory, taking the heaviest doubly magic nucleus $^{208}$Pb as an example, we firstly evaluate the propagated uncertainties of universal potential parameters for three typical types of single-particle energies in the phenomenological Woods-Saxon mean field. Accepting the Woods-Saxon modeling with uncorrelated model parameters, we find that the standard deviations of single-particle energies obtained by the Monte Carlo simulation and the error propagation rules are in good agreement with each other. It seems that the energy uncertaintis of the single-particle levels  regularly evoluate with some quantum numbers to a large extent for the given parameter uncertainties. Further, the correlation properties of the single-particle levels within the domain of input parameter uncertainties are analyzed using the method of statistical analysis, e.g., with the aid of Pearson correlation coefficients. It is found that the positive, negative or unrelated relationship may appear between two selected single-particle levels, which will be very helpful for evaluating the theoretical uncertainty related to the single-particle levels (e.g., $K$ isomer) in nuclear structural calculations.

\end{abstract}

\maketitle

%============================================================================================================================================%
\section{Introduction}
The fundamental theory of the strong interactions is quantum chromodynamics \cite{Wilczek1982}. As a final goal, all the phenomena in nuclear structure are expected to be derived from the interactions of quarks and gluons. To date, however, such a goal remains a daunting one to an extent though the density functional theory is attempting to approach it. In practice, in order to make the task tractable and more physically intuitive, numerous simplifications are usually made in theoretical modeling for nuclei. As is well known, the first approximation is certainly the use of the concept of nucleons and their interactions, which has been adopted in nearly all contemporary theories of nuclear structure. Further, the mean-field approximations and nucleon effective interactions are respectively proposed due to the difficulty of sloving the many-body problem and the complexity of the nucleon-nucleon interactions. Generally speaking, theoretical models for nuclear structure can be grouped into $ab$ $initio$ methods, mean-field theories, shell model theories, etc (cf. Ref.~\cite{M.Bender2003} and references therein).

Nuclear mean-field theories include phenomenological or empirical \cite{M. L. Gorelik2012,S G.Nilsson1969,Dudek1980,S.Cwiok1987} (e.g. the nuclear potentials of Woods-Saxon and Nilsson types) and self-consistent \cite{S.Goriely20091,S.Goriely2009,M.Bender1999} (e.g. numerous variants related to the Hartree-Fock approximation) ones, which assume that all the nucleons independently move along their orbits. In this type of nuclear theories, the underlying element contributing to good quality for theoretical calculations is the reliable mean-field single-particle energies, which sensitively depend on the corresponding Hamiltonian modeling and model parameters. For a defined mathematical model, the sampling (selection) and the quality of the experimental data will determine the resulting optimal parameter set and its quality. In principle, this can be done through the standard statistical fitting procedures, such as the least squares and $\chi^2$ fitting \cite{A.Savitzky1964,E.B.Wilson1931,K. Levenberg1944,Djelloul2020}. Then the physical quantity can be computed using the `optimal paremeters'. However, in the language of statistics, the overfitting (underfitting) phenomenon may appear if the model contains more (less) parameters. For instance, it was pointed that the so-called realistic model-interactions appear most of the time strongly over-parameterised \cite{Irene2018}. Therefore, there will remain uncertainties orginating from the size of sample database, the errors of the experimental data, the limited reliability of the model and the numerical method \cite{JDobaczewski2014}. In recent years, model prediction capacities and estimations of theoretical uncertainties are stongly interested in many subfields of physics and technological applications\cite{I.Dedes2019,Hv2016,Yuan2016,wangningcpc,Reinhard2013}. Even, it was pointed out that model predictions without properly quantified theoretical errors will be of very limited utility~\cite{J Piekarewicz2015}.

The phenomenological mean field, e.g., the realistic Woods-Saxon potential, has been used for many decades in nuclear physics and is considered as of very high predictive power for single-nucleon energies whereas related computing algorithms remain relatively simple. The model uncertainties and predictive power of spherically symmetric Woods-Saxon mean field have been investigated \cite{Irene2018}, paying particular attention to issues of the parameter adjustment and parametric correlations. Prior to this work, we, based on the one-body Hamiltonian with a phenomenological mean field of deformed Woods-Saxon type, have performed some studies~\cite{chaicpc,prcyang,li2019,chai2018} for different isotopes within the framework of macroscopic-microscopic model \cite{moller2008,W.Nazarewicz1989} and cranking approximation \cite{M.J.A1983,F.R.Xu2000}, focusing on different ground-state and high-spin nuclear properties. The main interest of our present work is not the fitting of the new parameters, the parameter uncertainties and the investigation of parameter correlations but rather studying the propagation of the reasonably given parameter uncertainties and the statistical correlation properties of the calculated single-particle levels within the domain of input parameter uncertainties using the same Woods-Saxon Hamiltonian. So far, such a systematical study is scarce and meaningful, especially for the theoretical calculations (e.g., $K$ isomer predictions) depended strongly on single-particle levels. As is well known, the single-particle levels are independent (which means the eigenfunctions of the Hamiltonian operator are orthogonal for different levels) in the mean-field approximation without the inclusion of the residual interaction. The wording of the `correlation properties' for the levels may be considered to be not suitable, even be seriously misunderstood by a general reader. Therefore, it should be particularly noted that the correlation property mentioned here means the statistical correlation (rather than something else, e.g., the correlation between the spin partners $j=l\pm1/2$) used for revealing the linear relationship of any two levels within the small domains related to their energy uncertainties. The calculated single-particle levels have the nature of the probability distribution (namely, the property of a stochastic quantity) after considering the uncertainty propagation of model parameters. That is to say, each calculated single-particle level will have the fixed value when calculating at a fixed mean field without the consideration of the model parameter uncertainties, whereas it will possess a stochastic value near its `fixed' one once the model parameter uncertainties are considered. In present work, as one of the aims, we will investigate the correlations between these stochastic values rather than the `relationships' of those `fixed' ones. It should also be noted that we will, first of all, accept the Woods-Saxon modeling with independent model parameters and then take the doubly magic nucleus $^{208}$Pb (which has always been regarded as a benchmark in the study of nuclear structure) as an example to perform the present investigation. The parameter uncertainties for the Woods-Saxon potential, even parameter correlations, have been estimated based on the maximum likelihood and the Monte Carlo methods~\cite{Irene2018, wangningcpc}.

The paper is organized in the following way. In Sec. II, by three subsections, we briefly introduce our theoretical framework on single-particle Hamiltonian, Monte Carlo method, propagation of the uncertainty and pearson product moment correlation. Four subsections of Sec. III present our results and discussion on the evaluation of universal potential parameters, generating pseudo data, uncertainties of single-particle energies and correlation effects between them. Finally, we give a summary in Sec. IV.
%============================================================================================================================================
\section{THEORETICAL FRAMEWORK}

%%%%%%%%%%%%%%%%%%%%%%%%%%%%%%%modle%%%%%%%%%%%%%%%%%%%%%%%%%%%%%%%%%%%%%%%%%%%%
%123456789 123456789 123456789 123456789 123456789 123456789 123456789 123456789
Given that our main goal are the uncertainty evolution of single-particle levels and the assessment of correlations among them due to the error propagation of model paremeters rather than the Hamiltonian modeling, the fitting of parameters or other physics issues. Here we will review some related points which are helpful for the general readers though there are numerous related references for each part.
%%%%%%%%%%%%%%%%%%%%%%%%%%%%%%%%%%%%%%%%%%%%%%%%%%%%%%%%%%%%%%%%%%%%%%%%%%%%%%%%
\subsection{Woods-Saxon single-particle Hamiltonian}

%%%%%%%%%%%%%%%%%%%%%%%%%%%%%%%%%%%%%%%%%%%%%%%%%%%%%%%%%%%%%%%%%%%%%%%%%%%%%%%%
%123456789 123456789 123456789 123456789 123456789 123456789 123456789 123456789

%%%%%%%%%%%%%%%%%%%%%%%%%%%%%%%%%%%%%%%%%%%%%%%%%%%%%%%%%%%%%%%%%%%%%%%%%%%%%%%%
%123456789 123456789 123456789 123456789 123456789 123456789 123456789 123456789
The single-particle levels and wave functions are calculated by solving numerically the stationary Schr\"{o}dinger equation with an average nuclear field of Woods-Saxon type.  The single-particle Hamiltonian for this equation  is given by~\cite{Dudek1980,S.Cwiok1987}
%%%%%%%%%%%%%%%%%%%%%%%%%%%%%%%%%%%%%%%%%%%%%%%%%%%%%%%%%%%%%%%%%%%%%%%%%%%%%%%%
%123456789 123456789 123456789 123456789 123456789 123456789 123456789 123456789
\begin{eqnarray}
    H_{\rm WS}
    &=&
    -\frac{\hbar^2}{2m}\nabla^2
    +
    V_{\rm cent}(\vec{r};\hat{\beta})
    +
    V_{\rm so}(\vec{r},\vec{p},\vec{s};\hat{\beta})
    \nonumber\\
    &&+
    \frac{1}{2}(1+\tau_3)V_{\rm Coul}(\vec{r},\hat{\beta}),
    \nonumber\\
                                                                  \label{eqn.01}
\end{eqnarray}
%%%%%%%%%%%%%%%%%%%%%%%%%%%%%%%%%%%%%%%%%%%%%%%%%%%%%%%%%%%%%%%%%%%%%%%%%%%%%%%%
%123456789 123456789 123456789 123456789 123456789 123456789 123456789 123456789
where the Coulomb potential $V_{Coul}(\vec{r},\hat{\beta})$ defined as a
classical electrostatic potential of a uniformly charged drop is added for
protons. The first part in the right side of Eq.~(\ref{eqn.01}) is the kinetic energy term. The central part of the Woods-Saxon potential which control mainly the number
of levels in the potential well is \cite{S.Cwiok1987}
%%%%%%%%%%%%%%%%%%%%%%%%%%%%%%%%%%%%%%%%%%%%%%%%%%%%%%%%%%%%%%%%%%%%%%%%%%%%%%%%
%123456789 123456789 123456789 123456789 123456789 123456789 123456789 123456789
\begin{equation}
    V_{\rm cent}(\vec{r},\hat{\beta})
    =
    \frac{V_0[1\pm\kappa(N-Z)/(N+Z)]}{1+{\rm exp[dist}_\Sigma(\vec{r},\hat{\beta})/a]},
                                                                  \label{eqn.02}
\end{equation}
%%%%%%%%%%%%%%%%%%%%%%%%%%%%%%%%%%%%%%%%%%%%%%%%%%%%%%%%%%%%%%%%%%%%%%%%%%%%%%%%
%123456789 123456789 123456789 123456789 123456789 123456789 123456789 123456789
where the plus and minus signs hold for protons and neutrons, respectively and
the $a$ is the diffuseness parameter of the nuclear surface.
%%%%%%%%%%%%%%%%%%%%%%%%%%%%%%%%%%%%%%%%%%%%%%%%%%%%%%%%%%%%%%%%%%%%%%%%%%%%%%%%
%123456789 123456789 123456789 123456789 123456789 123456789 123456789 123456789
The spin-orbit potential, which can strongly affects the level order, is defined
by
%%%%%%%%%%%%%%%%%%%%%%%%%%%%%%%%%%%%%%%%%%%%%%%%%%%%%%%%%%%%%%%%%%%%%%%%%%%%%%%%
%123456789 123456789 123456789 123456789 123456789 123456789 123456789 123456789
\begin{eqnarray}
    V_{\rm so}(\vec{r},\vec{p},\vec{s};\hat{\beta})
    &=&
    -\lambda
    \Big[
    \frac{\hbar}{2mc}
    \Big]^2
    \nonumber\\&&
    \bigg \{
    \nabla\frac{V_0[1\pm\kappa(N-Z)/(N+Z)]}{1
    +
    \textrm{exp}[\textrm{dist}_{\Sigma_{\rm{so}}}
    (\vec{r},\hat{\beta})/a_{\rm{so}}]}
    \bigg \}
    \times\vec{p}\cdot\vec{s},
    \nonumber\\
                                                                  \label{eqn.03}
\end{eqnarray}
%%%%%%%%%%%%%%%%%%%%%%%%%%%%%%%%%%%%%%%%%%%%%%%%%%%%%%%%%%%%%%%%%%%%%%%%%%%%%%%%
%123456789 123456789 123456789 123456789 123456789 123456789 123456789 123456789
where $\lambda$ denotes the strength parameter of the effective spin-orbit force
acting on the individual nucleons. In Eq.~(\ref{eqn.02}), the term
${\rm dis}_\Sigma(\vec{r},\hat{\beta})$ indicates the distance of a point $\vec{r}$
from the nuclear surface $\Sigma$. The nuclear surface is parametrized in terms of the multipole
expansion of spherical harmonics ${\rm Y}_{\lambda\mu}(\theta,\phi)$, namely,
%%%%%%%%%%%%%%%%%%%%%%%%%%%%%%%%%%%%%%%%%%%%%%%%%%%%%%%%%%%%%%%%%%%%%%%%%%%%%%%%
%123456789 123456789 123456789 123456789 123456789 123456789 123456789 123456789
\begin{equation}
    \Sigma:
    R(\theta,\phi)
    =
    r_0A^{1/3}c(\hat{\beta})
    \Big[
    1
    +
    \sum_{\lambda}
    \sum_{\mu=-\lambda}^{+\lambda}
    \alpha_{\lambda\mu}
    {\rm Y}^*_{\lambda\mu}(\theta,\phi)
    \Big],
                                                                  \label{eqn.04}
\end{equation}
%%%%%%%%%%%%%%%%%%%%%%%%%%%%%%%%%%%%%%%%%%%%%%%%%%%%%%%%%%%%%%%%%%%%%%%%%%%%%%%%
%123456789 123456789 123456789 123456789 123456789 123456789 123456789 123456789
where the function $c(\hat{\beta})$ ensures the conservation of the nuclear
volume with a change in the nuclear shape and $\hat{\beta}$ denotes the set of
all the considered deformation parameters. It is similar in Eq.~\ref{eqn.03} but the new surface $\Sigma_{so}$ needs to calculate using the different radius parameter.

Based on the Woods-Saxon Hamiltonian as mentioned above, the Hamiltonian matrix is calculated by using the axially deformed harmonic-oscillator basis in the cylindrical coordinate system with the principal quantum number $N\leqslant 12$ and 14 or protons and neutrons, respectively. Then, after a diagonalization procedure, the single-particle levels and their wave functions can be obtained. It is shown that the calculated single-particle levels with such a basis cutoff will be sufficiently stable with respect to a possibly basis enlargement in present work. Of course, one can see that for a given $(Z,N)$ nucleus, the calculated energy levels $\{e\}_{\pi,\nu}$ depend on two sets of six free parameters,
\begin{equation}
\{V^c,r^c,a^c,\lambda^{so},r^{so},a^{so} \}_{\pi,\nu},
                                                                  \label{eqn.05}
\end{equation}
one set with the symbol $\pi$ for protons, and the other set with $\nu$ for neutrons; the superscripts `c' and `so' denote the abbreviations for `central' and `spin-orbit', respectively.

For convenience, we define the parameter set $\{p\}\equiv \{p_1,p_2,p_3,p_4,p_5,p_6\}$, which is associated to the original one as follows,
\begin{equation}
  \{p_1,p_2,p_3,p_4,p_5,p_6\}_{\pi,\nu} \rightarrow \{V^c,r^c,a^c,\lambda^{so},r^{so},a^{so}\}_{\pi,\nu}.
                                                                  \label{eqn.06}
\end{equation}
Further, following the notation of Ref. \cite{F.J.F2011,R.P.G2010}, we can denote a point in such a parameter space by $\mathrm{p} =(p_1,p_2,p_3,p_4,p_5,p_6)$.
According to the inverse problem theory, the model parameters are usually determined by fitting to a set of observables within a selected sample (e.g., the available sample database of the experimental single-particle levels). For a given mathematical modeling, e.g. accepting the Woods-Saxon Hamiltonian with free parameters, the optimum parametrization $\mathrm{p}^{\rm o}$ can usually be obtained by a least-squares fit with the global quality measure \cite{R.P.G2010,F.J.F2012,Philip.R2003},
\begin{equation}
\chi^2(\mathrm{p})
=
\sum_{n=1}^{N}\left(
\frac{O_n^{({\rm th})}(\mathrm{p})
-O_n^{({\rm exp})}
}{\Delta O_n}
\right)^2
                                                                  \label{eqn.07}
\end{equation}
where `th' stands for the calculated values, `exp' for experimental data and $\Delta O$ for adopted errors, which generally contain the contributions from both experimental and theoretical aspects. Note that the definition of the objective function $\chi^2$ is standard and several powerful techniques for finding its minimum value have already been developed. The universal parameter set used in present investigation is, indeed, such one `optimal' parameter set. Having determined $\mathrm{p}^{\rm o}$, in principle, any physics quantity, e.g, the single-particle level $e_i$, can be computed at $e_i(\mathrm{p}^{\rm o})$. From this point, we can to an extent regard the calculated energy level $e_i$ as the function of the corresponding parameter set $\{p\}$, namely,
\begin{equation}
e_{i}=e_{i}(p_1,p_2,p_3,p_4,p_5,p_6).
                                                                  \label{eqn.08}
\end{equation}
There is no doubt, the $\mathrm{p}$ value depends on the size and quaility of the selected sample database. In fact, the functional relationship of Eq.~(\ref{eqn.08}) expresses not only a physcial law but also the measured and calculated processes.  All the uncertainties during the physical modeling, the experimental measurement and the theoretical calculation may lead to the uncertainty of the $\mathrm{p}$ value. On the contrary, the uncertainty of the $\mathrm{p}$ value will propagate to the results during the calculations.
%%%%%%%%%%%%%%%%%%%%%%%%%%%%%%%%%%%%%%%%%%%%%%%%%%%%%%%%%%%%%%%%%%%%%%%%%%%%%%%%
%123456789 123456789 123456789 123456789 123456789 123456789 123456789 123456789

%%%%%%%%%%%%%%%%%%%%%%%%%%%%%%%%%%%%%%%%%%%%%%%%%%%%%%%%%%%%%%%%%%%%%%%%%%%%%%%%
\subsection{Uncertainty estimation of single-particle levels}
By reasonably assuming that input parameters $\{p \}$ are Gaussian random variables, we will be able to estimate the uncertainties of the single-particle levels due to input uncertainties of potential parameters using conventional analysis method \cite{Rochman.D2011,Rochman.D2014,Sciacchitano.A2016} (e.g., the formula of uncertainty propagation) and Monte Carlo method \cite{N.Metropolis1949,J.M.H1965,Lang.G.H1993,D.J.Dean1999,S.E.Koonin1997}. Based on the functional relationship of Eq.~(\ref{eqn.08}) and the uncertainty propagation formula, the uncertainty of the $i$th single-particle level $e_i$ with random and uncorrelated inputs can be given analytically by
\begin{equation}
\sigma_{e_{i}}=\sqrt{\sum_{j=1}^6\left(\frac{\partial e_{i}}{\partial p_j}\right)^2\cdot \sigma_{p_j}^2},
                                                                  \label{eqn.9}
\end{equation}
where $\sigma_{p_j}$ is the standard deviation of the input parameter $p_j$; the partial derivative $\partial e_i/\partial p_j$ is usually called sensitivity coefficient, which gives the effect of the corresponding input parameter on the final result. Note that both linearity of the function (at least, near the calculated point $p_j$) and `small' uncertainty of input parameter are prerequisites of the conventional method of uncertainty estimation. However, there is no such limitation for Monte Carlo simulation method, which can handle both small and large uncertainties in the input quantities. Moreover, the Monte Carlo simulation, which can be generally defined as the process of replication of the `real' world, has the ability to take account of partial correlation effects for input parameters. It is also convenient to study the correlation effect e.g. between two Gaussian-distributed variables whereas the conventional method cannot do this.

As known in such a simulation, the availability of high quality Gaussian random numbers is of importance. Generally speaking, the realization of Gaussian-random-number generator can adopt the software and the hardware methods. The former has the limited speed and poor real-time characteristic while the latter (which is based on digital devices) is not only fast, real-time, but also has good flexibility and accuracy. At present, the majority of the frequently used digital methods for generating Gaussian random variables are based on transformations from uniform random variables. Popular methods, for instance, include the Ziggurat method \cite{G.Marsaglia2000}, the inversion method \cite{W.H2003}, the Wallace method \cite{C.Wallace1996}, the Box-Muller method \cite{G.Box1958,Lee2006,A.Alimohammad2008,Emmanuel B2003} and so on. In present work, we realize the hardware Gaussian random number generators using the Box-Muller algorithm.
Namely, taking each value $p_j^o$ of the universal parameter set $\{p_1^o,p_2^o,p_3^o,p_4^o,p_5^o,p_6^o\}$ as the corresponding mean value, one can generate the random and uncorrelated input parameter $p_j$ following a normal distribution $N(p_j^o,\sigma_{p_j})$. With a large sample of input parameters, the uncertainties of single-particle levels can be estimated. For instance, considering the uncertainty of one input parameter $p_j$ and keeping other universal values unchanged, the variance of the calculated $e_i$ can be given by
\begin{equation}
\sigma_{e_{i}}^2=\frac{1}{N-1}\sum_{k=1}^N [e_{i}(p_{jk})-e_{i}(p_{j}^o)]^2,
                                                                  \label{eqn.10}
\end{equation}
where the sampling number $N$ should be chosen to be sufficiently large (e.g., 10 000 or more). Similar calculations can be performed when the uncertainties of two or more input parameters are opened. Therefore, we will be able to investigate the effects of the uncertainties of different input parameters and their combinations on the uncertainties of single-particle levels.

%%%%%%%%%%%%%%%%%%%%%%%%%%%%%%%%%%%%%%%%%%%%%%%%%%%%%%%%%%%%%%%%%%%%%%%%%%%%%%%%
\subsection{Pearson Product Moment Correlation}

The single-particle levels with certain uncertainties can usually be regarded as the input variables in further nuclear mean-field calculations, e.g., the $K$ isomeric calculations. In this case, for evaluating the further theoretical predictions, it will be very useful to know both the uncertainties and correlation properties of the single-particle levels. As a simple example, the energy uncertainty of one-particle one-hole ($1p1h$) excitation will directly relate to two corresponding single-particle levels to a large extent. Once we can arbitrarily regard the excitation energy $E_{1p1h}$ as the function of two single-particle levels $e_1$ and $e_2$ with the standard deviations $\sigma_{e_1}$ and $\sigma_{e_2}$, respectively, namely,
\begin{equation}
E_{1p1h}^*=f(e_1,e_2).
                                                                  \label{eqn.11}
\end{equation}
Regardless of whether or not $e_1$ and $e_2$ are independent, the standard uncertainty of such an excited state can be written as~\cite{Papadopoulos2001}
\begin{eqnarray}
&&\sigma_{E_{1p1h}^*} \nonumber \\
&=&\sqrt{\left(\frac{\partial f}{\partial e_1}\right)^2 \sigma_{e_1}^2+\left(\frac{\partial f}{\partial e_2}\right)^2 \sigma_{e_2}^2+2\frac{\partial f}{\partial e_1}\frac{\partial f}{\partial e_2}\rho(e_1,e_2)\sigma_{e_1}\sigma_{e_2}}, \nonumber \\
                                                                  \label{eqn.12}
\end{eqnarray}
where the quantity $\rho(e_1,e_2)$ is the Pearson correlation coefficient, which is given by \cite{J.Lee.R1988,K.Pearson1909}
\begin{equation}
\rho(e_1,e_2)=\frac{cov(e_1,e_2)}{\sigma_{e_1}\sigma_{e_2}}.
                                                                  \label{eqn.13}
\end{equation}
Such a cross-correlation coefficient measures the strength and direction of a linear relationship between two variables, e.g., $e_1$ and $e_2$. The greater the absolute value of the correlation coefficient, the stronger the relationship. The extreme values of -1 and 1 indicate a perfectly linear relationship where a change in one variable is accompanied by a perfectly consistent change in the other. At these two cases,
all of the data points fall on a line. A zero coefficient represents no linear relationship. That is, as one variable increases, there is no tendency in the other variable to either increase or decrease. When the cross-correlation coefficient is in-between 0 and +1/-1, there will be a relationship, but all the points don’t fall on a line. The sign of the correlation coefficient represents the direction of the linear relationship. Positive coefficients indicate that when the value of one variable increases, the value of the other variable also tends to increase. Positive relationships produce an upward slope on a scatterplot. Negative coefficients represent the cases: when the value of one variable increases, the value of the other variable tends to decrease. Correspondingly, negative relationships produce a downward slope. It should be noted that Pearson’s correlation coefficient, which measures only linear relationship between two variables, will not detect a curvilinear relationship. For instance, when the scatterplot of two variables shows a symmetric distribution, the relationship may exist but the correlation coefficient is zero.

%%%%%%%%%%%%%%%%%%%%%%%%%%%%%%%%%%%%%%%%%%%%%%%%%%%%%%%%%%%%%%%%%%%%%%%%%%%%%%%%
%%%%%%%%%%%%%%%%%%%%%%%%%%%%%%%%%%%%%%%%%%%%%%%%%%%%%%%%%%%%%%%%%%%%%%%%%%%%%%%%
\section{RESULTS AND DISCUSSION}

%%%%%%%%%%%%%%%%%%%%%%%%%%%%%%%%%%%%%%%%%%%%%%%%%%%%%%%%%%%%%%%%%%%%%%%%%%%%%%%%
\subsection{Evaluation of Woods-Saxon potential parameters}

\begin{table*}
\begin{minipage}{16cm}
\begin{center}
\caption{
  Various parameter sets for Woods-Saxon potential
                                                                  \label{table}
}
\begin{tabular}{p{2.4cm} ccccccccc p{1.4cm}  ccccccc}
\hline \hline\small
        {Parameter}   &   & {$V_{0}$(MeV)}         &{$r^{c}_{0}$(fm)}
        &{$a^{c}_{0}$(fm)}   &  {$\lambda$} &  {$r^{so}_{0}$(fm)}  &  {$a^{so}_{0}$(fm)}
                                                                              \\
\hline
              Wahlborn \cite{J.Blomqvist1960}  &  &51.0      &1.27    &0.67  &32.0   &1.27   &0.67     \\
\hline
              Rost \cite{E.Rost1968}          &n  &49.6      &1.347     &0.7   &31.5   &1.280  &0.7  \\
                                              &p  &          & 1.275    &      &17.8   &0.932  &        \\
\hline
              Chepurnov \cite{V.A.Chepurnov1967}   &  &53.3       &1.24    &0.63  &$23.8\cdot(1+2I)$     &1.24   &0.63  \\

\hline
              New \cite{J.Dudek1979}          &n  &49.6       &1.347    &0.7   &$*$    &$*$  &0.7  \\
                                              &p  &           & 1.275   &      &$*$    &$*$  &     \\
\hline
              Universal \cite{S.Cwiok1987}     &n  &49.6       &1.347    &0.7   &35.0   &1.31  &0.7  \\
                                              &p  &           &1.275    &      &36.0   &1.32  &     \\
\hline
            Optimized \cite{J.Dudek1981}                 &n  &49.6     &1.347   &0.7  &36.0 &1.30   &0.7  \\
                                              &p  &          &1.275   &      &       &   &      \\
\hline
              Cranking \cite{Bhagwat.A2010,Meng2018}    &  &53.754     &1.19   &0.637  &29.494 &1.19   &0.637  \\
\hline \hline
\end{tabular}
\end{center}
\end{minipage}
\end{table*}
%%%%%%%%%%%%%%%%%%%%%%%%%%%%%%%%%%%%%%%%%%%%%%%%%%%%%%%%%%%%%%%%%%%%%%%%%%%%%%%%
%123456789 123456789 123456789 123456789 123456789 123456789 123456789 123456789
%description for Table. 1
In the phenomenological nuclear mean field, the realistic Woods-Saxon potential has shown certain advantages and is still used widely. For instance, it provides a good description of not only the ground-state properties but also the excited-state properties of nuclei. Nowadays, many authors are still working in different issues with the Woods-Saxon potential. Such a simple nuclear mean field has been successfully applied to explain and predict the nuclear equilibrium deformations, the hight-$K$ isomer, the nucleon binding energies, the fission barriers, numerious single-particle effects for superdeformed and fast rotating nuclei, and so on. As shown in Table~\ref{table}, there exist various parametrizations of the Woods-Saxon potential (cf. Ref.~\cite{S.Cwiok1987} and references therein), which are usually obtained by fitting the available single-particle data (or part of them, namely, one of the subdatabases) or other observables. Indeed, based on the same mathematical modeling and different sample databases/subdatabases, different parameter sets can be obtained. It can be seen that these parameter sets are somewhat different, even rather different for some quantities among them. Correspondingly, the different parameter sets are just suitable for a certain nuclear mass region. Sometimes, the difference of the corresponding quantity (e.g., single-particle energies) calculated theoretically using different parameter sets is referred to as model discrepancy, which can be evaluated by using different models and/or different parameter sets. The universal parameter set of the Woods-Saxon potential is one of the most commom parameter sets. In principle, it can be used for the `global' calculation for nuclei . In present paper, we will perform our investigation based on the universal parameter set.

%%%%%%%%%%%%%%%%%%%%%%%%%%%%%%%%%%%%%%%%%%%%%%%%%%%%%%%%%%%%%%%%%%%%%%%%%%%%%%%%
\begin{figure}[htbp]
\centering
\includegraphics[width=0.5\textwidth]{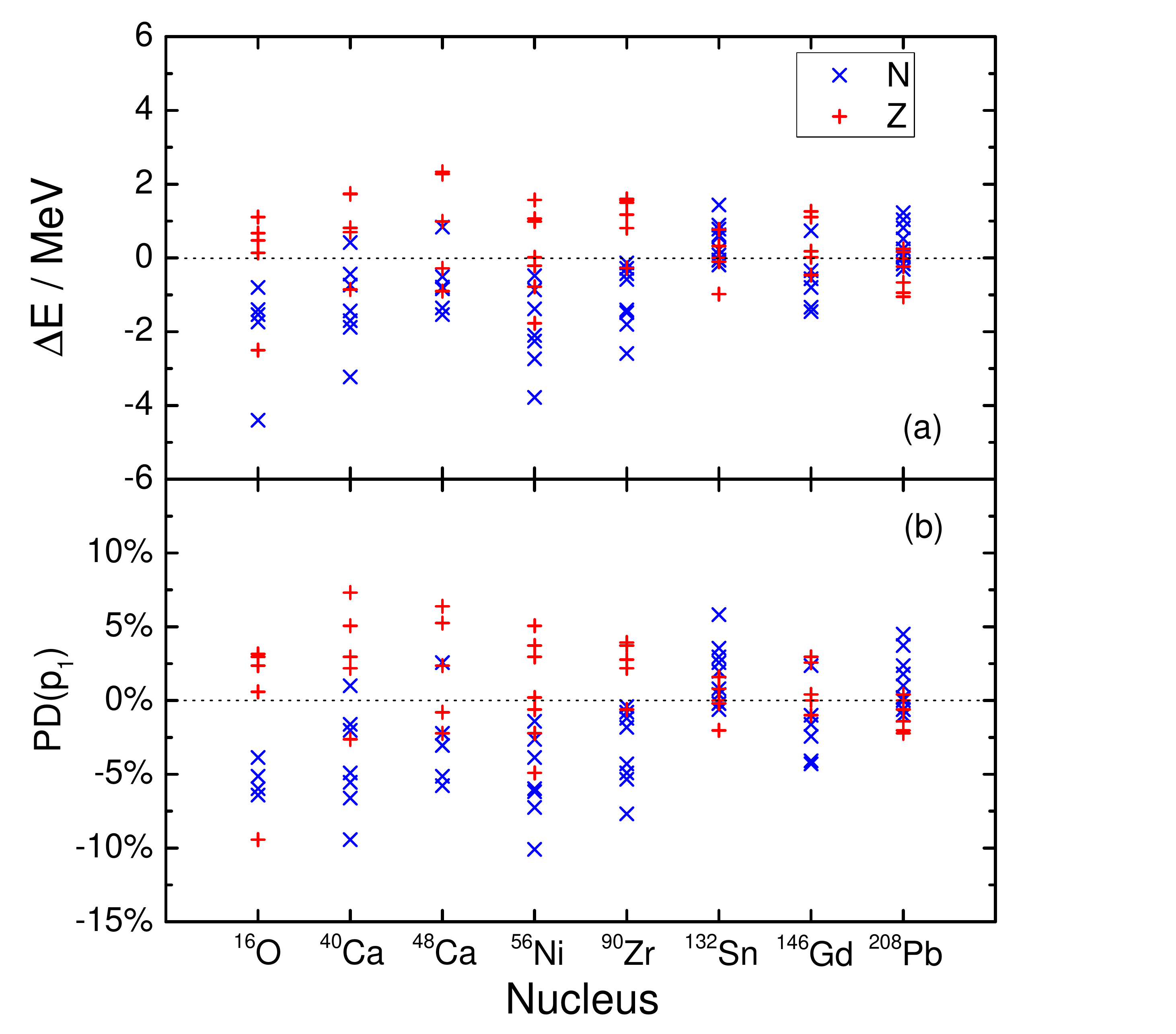}
\caption{
(Color online) (a) Discrepancies between the available experimental data and the calculated single-particle energies using the Woods-Saxon Universal parameter set for even-even nuclei $^{16}$O, $^{40}$Ca,$^{48}$Ca, $^{56}$Ni,$^{90}$Zr,$^{132}$Sn,$^{146}$Gd and $^{208}$Pb. The data are taken from Refs.~\cite{J.Dudek2010,N. Schwierz2007}. (b) Percentage differences between the `best' and the `optimal' $p_1$ (namely, $V_0$) parameters.  See text for more details.
                                                                    \label{fig1}
}
\end{figure}
%%%%%%%%%%%%%%%%%%%%%%%%%%%%%%%%%%%%%%%%%%%%%%%%%%%%%%%%%%%%%%%%%%%%%%%%%%%%%%%%
%123456789 123456789 123456789 123456789 123456789 123456789 123456789 123456789
%description for Fig.1
In order to evaluate the universal potential parameters, Figure.~\ref{fig1}(a) shows the discrepancies $\Delta E$ ($\equiv e_i^{\rm theo.}-e_i^{\rm exp.}$) of the calclated single-particle energies from the available data (e.g., eight spherical nuclei~\cite{N. Schwierz2007,J.Dudek2010}: $^{16}$O, $^{40}$Ca,$^{48}$Ca, $^{56}$Ni,$^{90}$Zr,$^{132}$Sn,$^{146}$Gd and $^{208}$Pb). The discrepancies show us that the single-particle levels generated by the universal parameters, in fact, cannot agree with the data very well (similar to the mass calculation~\cite{wudi2020}, the quest for some possibly missing interactions and `better' mathematic modeling will never stop). Moreover, most of the values are smaller or larger than the data. For instance, as can be seen, there exist the systematic overestimation and underestimation for protons and neutrons, respectively, especially in the lighter nuclei. To see the statistical properties of the parameters, the percentage difference, $PD(p_j)$, of the model parameter $p_j$ (as an example, $j=1$ here) extracted from the experimental data is presented in Fig.~\ref{fig1}(b), which is defined by
\begin{equation}
PD(p_j)=\frac{p_j^{\rm b}-p_j^{\rm o}}{\frac{p_j^{\rm b}+p_j^{\rm o}}{2}}\times 100\%,
                                                                  \label{eqn.14}
\end{equation}
where the $p_j^{\rm o}$ means the $j$th `optimal' (universal) value of $\{p\}$ parameters; the $p_j^{\rm b}$ parameter denotes the so-called `best' value which can be obtained based on the following method. For a certain model parameter, e.g. the potential depth $p_1$ (namely, $V_0$) of the Woods-Saxon parameters, we calculate the corresponding single-particle energies of a given nucleus by varying the value of this parameter $p_1$ around its optimal value $p_1^{\rm o}$ and keeping other parameters with universal values unchanged. If the discrepancy of the calculated single-particle energy for a certain nucleus from the corresponding experimental data equals zero, the ``best'' value $p_1^{\rm b}$ of this parameter $p_1$ for this nucleus is therefore obtained. In principle, for a large sample, we can extract the standard deviation $\sigma_{p_1}$ with $68.3\%$ confidence level for the parameter $p_1$. From Fig.~\ref{fig1}(b), it is found that the percentage differences distribute between $\pm 10\%$. There exist similar distributions for other potential parameters. Based on these statistical properties and some previous studies (cf. e.g, Refs.\cite{I.Dedes2019,wangningcpc}), we can evaluate the uncertainty of Woods-Saxon paramters to an extent. Further, we will be able to temporarily give the standard deviations $\{\sigma_{p}\} \equiv \{\sigma_{p_1},\sigma_{p_2},\sigma_{p_3},\sigma_{p_4},\sigma_{p_5},\sigma_{p_6}\}$ for the parameters $\{p\}$ within reasonable domains, taking the universal parameters $\{p_1^{\rm o},p_2^{\rm o},p_3^{\rm o},p_4^{\rm o},p_5^{\rm o},p_6^{\rm o} \}$ as the corresponding mean values.

\subsection{Producing pseudo-data of potential parameters}
%%%%%%%%%%%%%%%%%%%%%%%%%%%%%%%%%%%%%%%%%%%%%%%%%%%%%%%%%%%%%%%%%%%%%%%%%%%%%%%%

\begin{figure*}[htbp]
\centering
\includegraphics[height=0.7\textwidth]{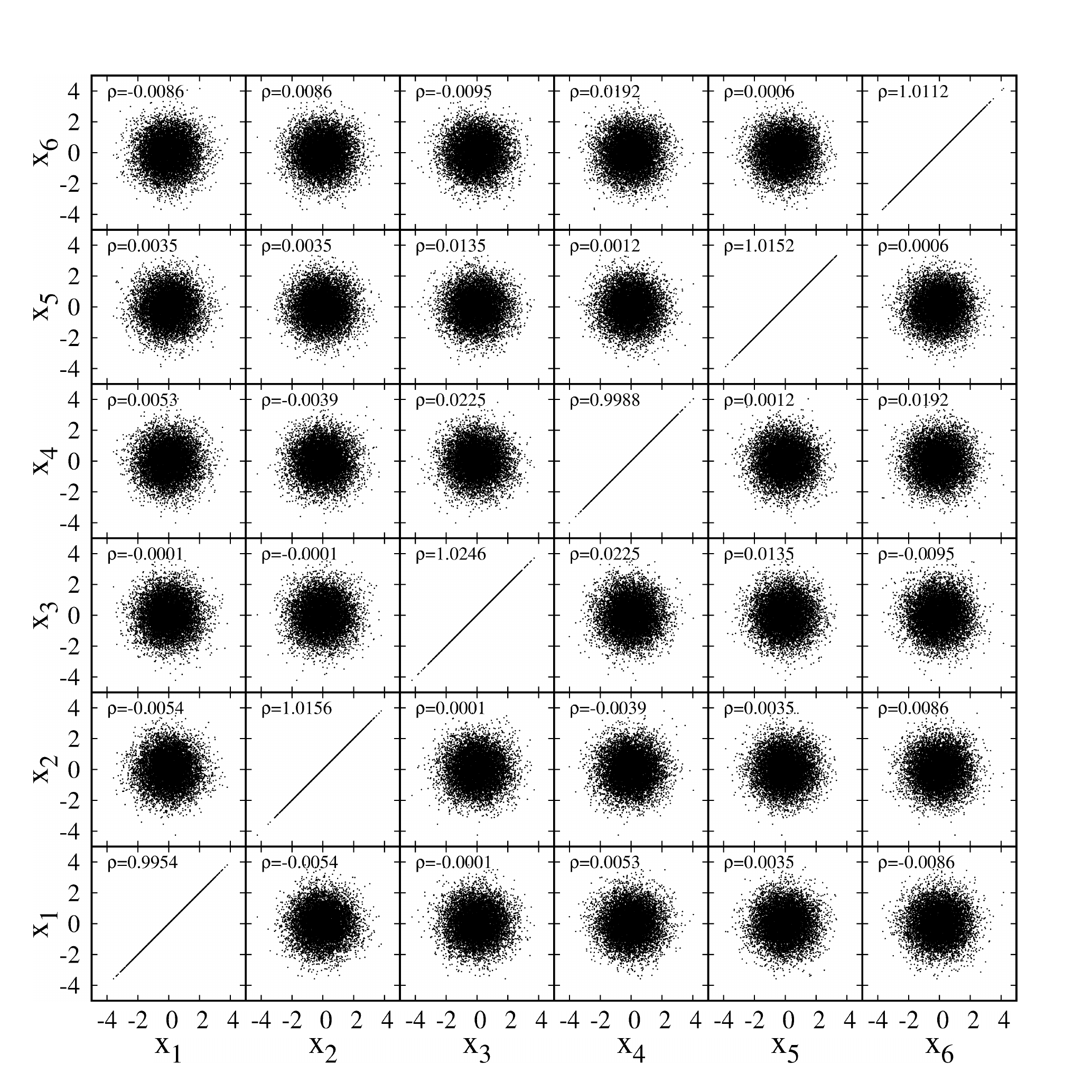}
\caption{
(Color online) Two-dimensional scatter plots, together with their corresponding correlation coefficients, between 6 independent WS model parameters.
                                                                   \label{fig2}
}
\end{figure*}
%%%%%%%%%%%%%%%%%%%%%%%%%%%%%%%%%%%%%%%%%%%%%%%%%%%%%%%%%%%%%%%%%%%%%%%%%%%%%%%%
%123456789 123456789 123456789 123456789 123456789 123456789 123456789 123456789
%description for Fig.2
Based on the given mean values $\{p^{\rm o}\}$ and the corresponding standard deviations $\{\sigma_p\}$, the Gaussian-distributed random sets $\{p \}$ can, in principle, be numerically generated in the spirit of the Monte Carlo approach. Considering the uncertainty estimations of the Woods-Saxon parameters and the sensitivity coefficients of single-particle levels, we, in practice, use a set of percentage coefficients $\{{\rm c}\}\equiv \{{\rm c}_1, {\rm c}_2, {\rm c}_3,{\rm c}_4, {\rm c}_5,{\rm c}_6\}= \{0.1\%,0.1\%,1\%,3\%,1\%,10\%\}$ to calibrate the standard deviations $\{\sigma_p\}$ during the calculations. That is, the standard deviations are given by
\begin{equation}
\begin{pmatrix} \sigma_{p_1}  \\ \sigma_{p_2} \\ \sigma_{p_3}  \\ \sigma_{p_4}  \\ \sigma_{p_5}  \\ \sigma_{p_6} \end{pmatrix}
=
\begin{pmatrix} c_1 & c_2 &c_3 &c_4 &c_5 & c_6 \end{pmatrix}
\begin{pmatrix} p_1^{\rm o}  \\ p_2^{\rm o} \\ p_3^{\rm o}  \\ p_4^{\rm o}  \\ p_5^{\rm o}  \\ p_6^{\rm o} \end{pmatrix}.
                                                                  \label{eqn.15}
\end{equation}
Such a set $\{\sigma_p\}$ may deviate from the `true' values to an extent but does not affect the conclusion of our investigation since they lie in the reasonable domains. Moreover, the strong overlaps of the `peaks' of single-particle levels can be avoided (as seen below). We perform the Woods-Saxon single-particle-level calculations with 10 000 samples for $\{p \}$, which is large engough to suppress the error coming from stochastic choices. To show the quality of the normally distrubited random quantities $\{p \}$, Figure~\ref{fig2} presents the two-dimensional scatter plots related to the 6 Woods-Saxon parameter sampling of neutrons, together with the corresponding correlation coefficients. Note that it is similar for protons. For comparison, the normal distribution $N(p_i^{\rm o}, \sigma_{p_i})$ is transformed into the standard normal distribution $N(0, 1)$ by defining the dimensionless parameter $x_i=(p_i-p_i^{\rm o})/\sigma_{p_i}$ in Fig.~\ref{fig2}. One can see the Gaussian-distributed and independent properties of these parameters. In addition, the calculated skewness and kurtosis values are zero, as expected, indicating the Gaussian-type distributions as well.

%%%%%%%%%%%%%%%%%%%%%%%%%%%%%%%%%%%%%%%%%%%%%%%%%%%%%%%%%%%%%%%%%%%%%%%%%%%%%%%%
\subsection{Uncertainties of single-particle energies}

With the sampling $\{p \}$, the uncertainties of the single-particle energies will be able to be exactly evaluated. Indeed, this is the advantage of the Monte Carlo method. For convenience, using the similar $\gamma-\gamma$ coincidence technique which is widely used for experimentally deducing the nuclear level scheme, we construct a level-level coincidence matrix (namely, a two dimensional histogram). Each axis of the matrix corresponds to the energy of the calculated single-particle levels. The matrix has a dimension of $4096 \times 4096$ channels, with the energy calibration $10$ keV/channel. Such a matrix gives an energy range from $-40.96$ to $0.00$ MeV, covering the range of the single-particle energies (e.g., all the bounded ones for neutrons) that we care for. By using the gated spectra on different level-level matrice, one can conveniently analyze the peak distributions of single-particle levels and even their correlation properties at different conditions.

%%%%%%%%%%%%%%%%%%%%%%%%%%%%%%%%%%%%%%%%%%%%%%%%%%%%%%%%%%%%%%%%%%%%%%%%%%%%%%%%
\begin{figure}[htbp]
\centering
\includegraphics[height=0.55\textwidth]{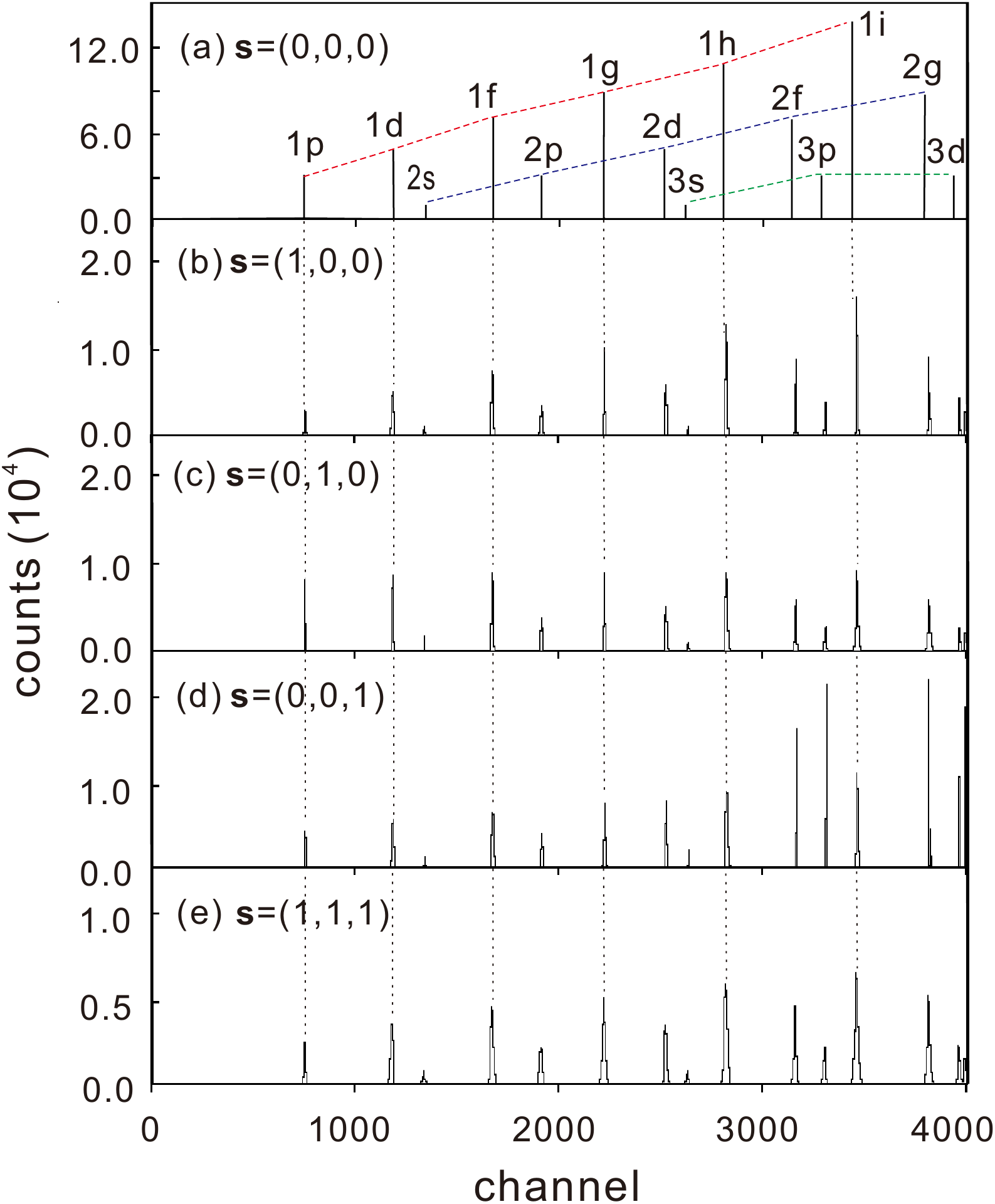}
\caption{
(Color online) Calculated single-neutron levels labeled as $\{nl\}$ in $^{208}$Pb (gated at the $1s$ level). The dotted lines  are provided to guide the eye. See the text for further calculated details.   
                               \label{fig3}
}
\end{figure}
%%%%%%%%%%%%%%%%%%%%%%%%%%%%%%%%%%%%%%%%%%%%%%%%%%%%%%%%%%%%%%%%%%%%%%%%%%%%%%%%
%123456789 123456789 123456789 123456789 123456789 123456789 123456789 123456789
%description for Fig.3
%%%%%%%%%%%%%%%%%%%%%%%%%%%%%%%%%%%%%%%%%%%%%%%%%%%%%%%%%%%%%%%%%%%%%%%%%%%%%%%%
\begin{figure}[htbp]
\centering
\includegraphics[width=0.45\textwidth]{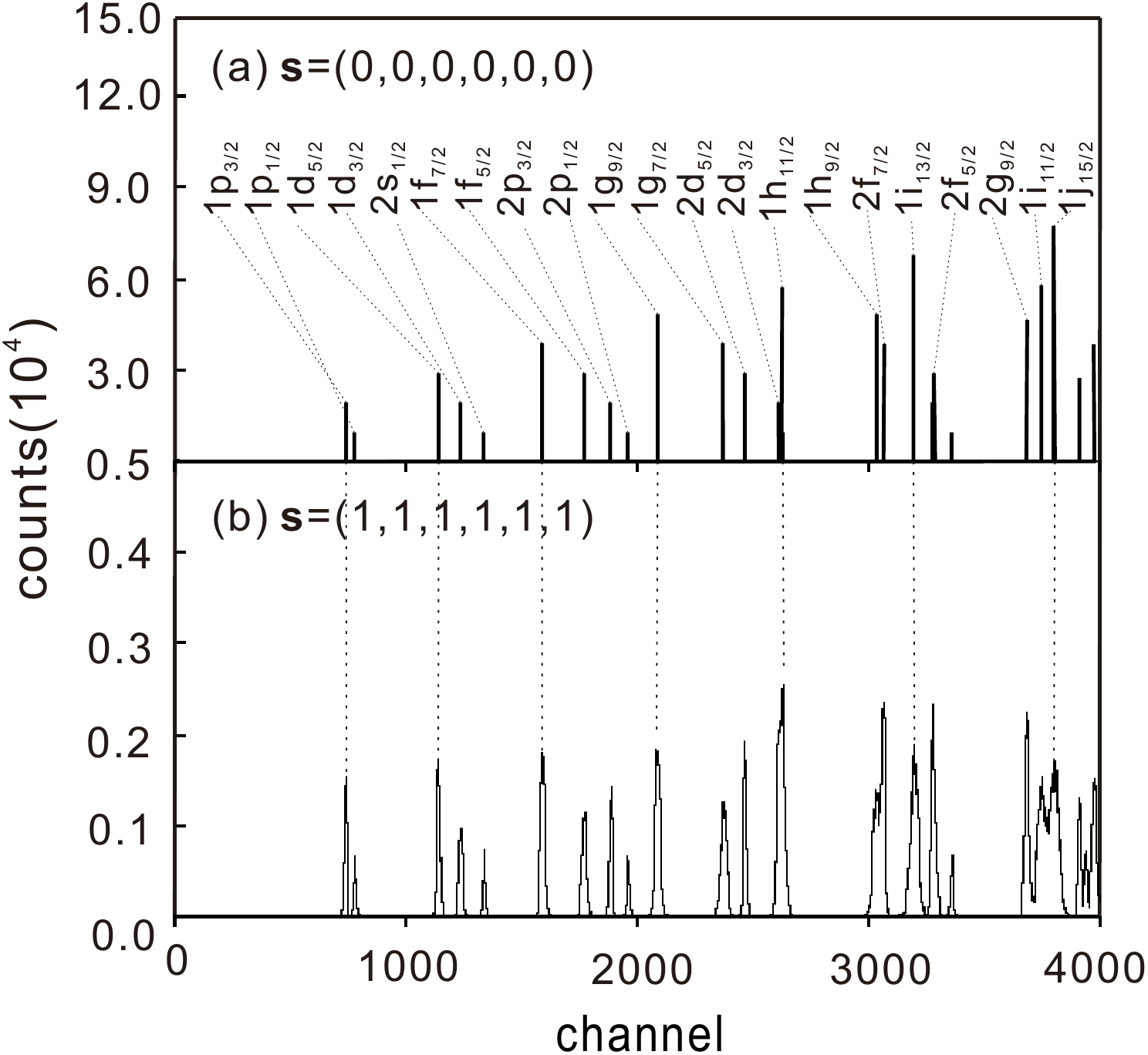}
\caption{
(Color online) Similar to Fig. \ref{fig3}, but the spin-orbit coupling is considered (gated at the $1s_{1/2}$ level). 
                                                                    \label{fig4}
}
\end{figure}

%%%%%%%%%%%%%%%%%%%%%%%%%%%%%%%%%%%%%%%%%%%%%%%%%%%%%%%%%%%%%%%%%%%%%%%%%%%%%%%%
\begin{figure}[hbp]
\centering
\includegraphics[width=0.45\textwidth]{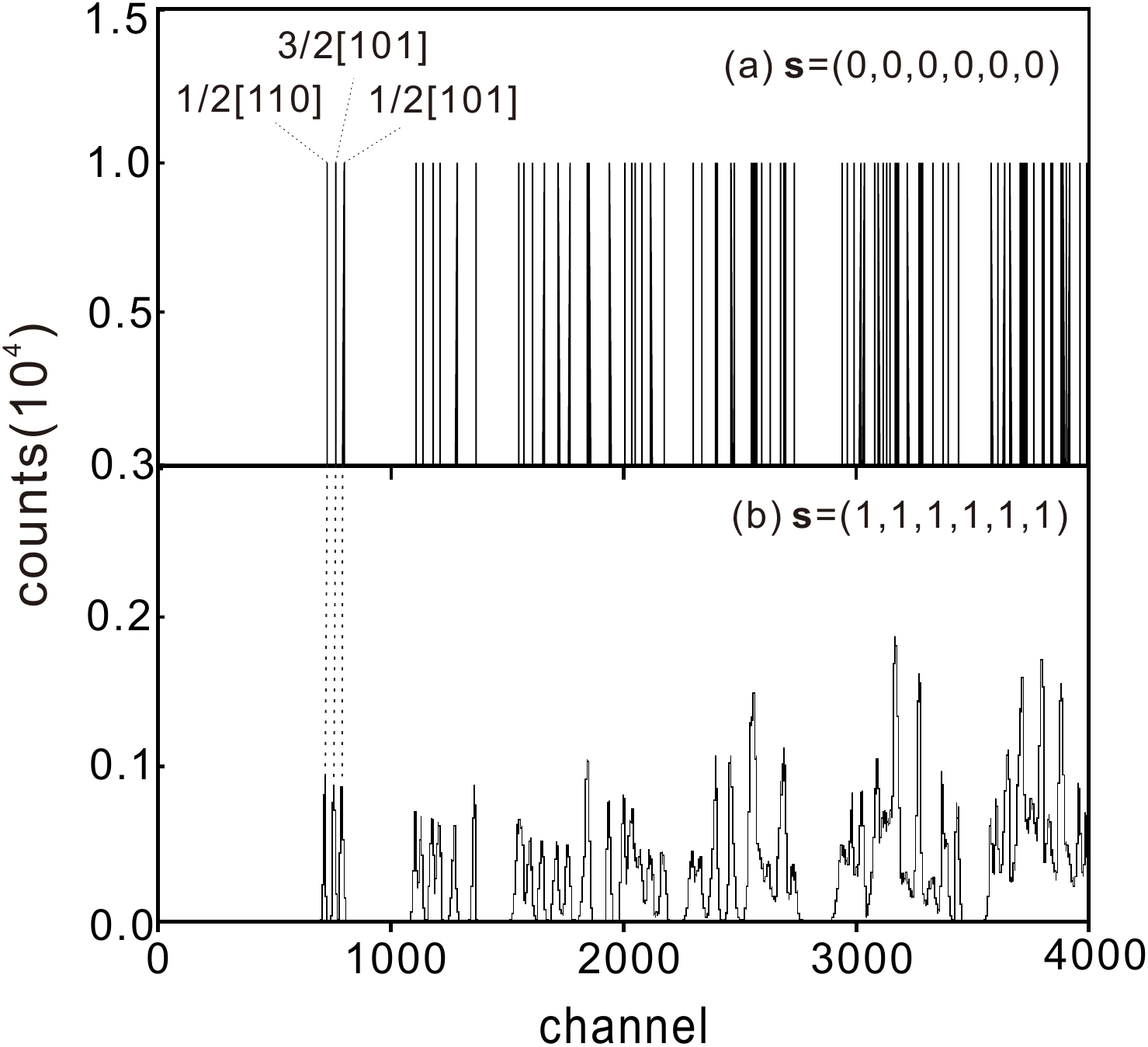}
\caption{
(Color online) Similar to Fig. \ref{fig4}, but the axail deformation is added (gated at the $1/2[100]$ level).
Note that only the first three $\{\Omega[Nn_z\Lambda]\}$ quantum numbers are given due to space limitations.
                                                                    \label{fig5}
}
\end{figure}
%%%%%%%%%%%%%%%%%%%%%%%%%%%%%%%%%%%%%%%%%%%%%%%%%%%%%%%%%%%%%%%%%%%%%%%%%%%%%%%%
%123456789 123456789 123456789 123456789 123456789 123456789 123456789 123456789
%description for Fig.6

As is well known, there are three typical types of single-particle levels during the evolution of nuclear models (for instance, from the harmonic oscillator model, adding strong spin-orbit coupling to obtain the shell model, and axial deformation to give the collective model). In this work, we use a more realistic Woods-Saxon potential (lies between the harmonic oscillator potential and the finite square well) to produce these three kinds of single-particle levels and study their energy uncertainties originated from model parameters. Similar to the parameter space $(p_1,p_2,p_3,p_4,p_5,p_6)$, let us define a correspondingly 6-dimensional `switch' space $(s_1,s_2,s_3,s_4,s_5,s_6)$, where $s_i=0$ or 1 (for $i=1, 2, \cdots,6$). Moreover, if $s_i=0$, the universal parameter $p_i^{\rm o}$ is always adopted (that is, the standard deviation $\sigma_{p_i}$ is not used). For $s_i=1$, it indicates that the sampling $p_i$ value is adopted (in other words, the parameter $\sigma_{p_i}$ is opened). Obviously, we can evaluate the effects of different parameter uncertainties and their combinations on single-particle levels by calculating at different points $\mathrm{s} =(s_1,s_2,s_3,s_4,s_5,s_6)$.

In Fig.3,  we show the spherical single-particle levels (labeled as $\{nl\}$ quantum numbers) calculated using the Woods-Saxon potential without the inclusion of the spin-orbit coupling. Note that in spectroscopic notation the bound states for angular momentum states with $l=0,1,2,3,4,5,\cdots$ are indicated with the letter $s,p,d,f,g,h,\cdots$, respectively. The projection spectra at different $\mathrm{s}$ points are obtained by gating at the $1s$ level.  Figure 4 shows the second kind of single-particle levels (labeled as $\{nlj\}$) calculated at two $\mathrm{s}$ points using the spherically Woods-Saxon potential with the spin-orbit part. In this case, the $l$ orbital will be splitted into two $j=l\pm \frac{1}{2}$ substates. Similar to Fig.4, in Fig. 5, we show the deformed Woods-Saxon single-particle levels (labeled as $\{\Omega[Nn_z\Lambda]\}$, the so-called Nilsson quantum numbers) calculated at $\beta_2=0.1$, an arbitrarily selected axial deformation value. In Fig.~\ref{fig5}, one can see that the peak heights of the deformed single-particle levels are all same, with the sampling value, 10 000, since the two-fold degenerate levels $\{\Omega[Nn_z\Lambda]\}$ are not degenerate any more. However, the levels labeled as $\{nl\}$ and $\{nlj\}$ have the $(2n+1)$- and $(j+\frac{1}{2})$-fold degeneracies, respectively, due to the spherical symmetry of the Woods-Saxon potential. As seen in Figs.~\ref{fig3}(a) and \ref{fig4}(a), the counts dividing by 10 000 indicate the degrees of degeneracy of the corresponding levels. Based on these gated spectra at different $\mathrm{s}$ points, we can analyze the distributed properties of the single-particle levels without the strong overlap. For instance, it is convenient to fit the distributions in Figs.3 and 4 while it is difficult in the right part of Fig.5 since the distributions overlap strongly.

%%%%%%%%%%%%%%%%%%%%%%%%%%%%%%%%%%%%%%%%%%%%%%%%%%%%%%%%%%%%%%%%%%%%%%%%%%%%%%%%
\begin{figure}[htbp]
\centering
\includegraphics[width=0.4\textwidth]{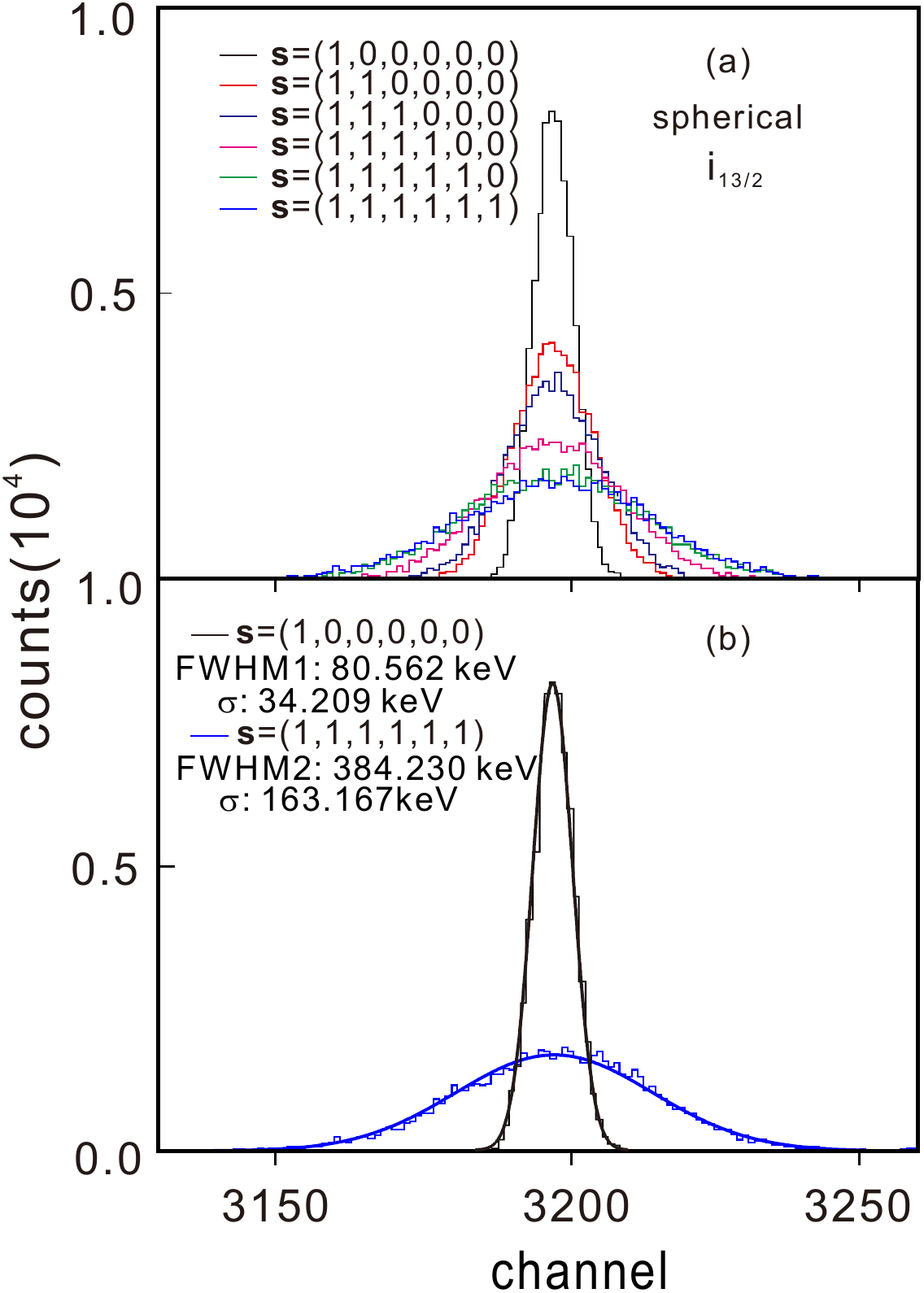}
\caption{
(Color online) (a) Distributions of the spherical $i_{13/2}$ level calculated at different $\mathrm{s} =(s_1,s_2,s_3,s_4,s_5,s_6)$ points. (b) The Gaussian fits to the distributions at $\mathrm{s} =(1, 0, 0, 0, 0,0)$ and $(1,1,1,1,1,1)$ points.
                                                                    \label{fig6}
}
\end{figure}
%%%%%%%%%%%%%%%%%%%%%%%%%%%%%%%%%%%%%%%%%%%%%%%%%%%%%%%%%%%%%%%%%%%%%%%%%%%%%%%%
%123456789 123456789 123456789 123456789 123456789 123456789 123456789 123456789
%description for Fig.6
In Fig.~\ref{fig6}(a), as an example, we show that the uncertainty evolution of the selected spherical $i_{13/2}$ level as more and more uncertainty parameters are opened.
It can be seen that the energy uncertainty of this level increases with increasing `$1$' in the `switch' space. The results of the Gaussian fits to the peaks at $\mathrm{s} =(1,0,0,0,0,0)$ and $(1,1,1,1,1,1)$ points are presented in Fig.~\ref{fig6}(b), including the standard deviations and the full width at half maximum (FWHM). The FWHM is a parameter commonly used to describe the width of a ``bump'', e.g., on a function curve which is given by the distance between points on the curve at which the function reaches half its maximum value. The FWHM can be used for describing the width of any distribution. For a normal distribution $N(\mu,\sigma)$, its FWHM is $2\sqrt{2{\rm ln2}}$ ($\sigma\approx 2.3548\sigma$). In principle, we can extract the standard deviation $\sigma_{e_i}$ for each single-particle level $e_i$ and further find the possible evoluation law. It is found that, in practice, the correct fitting will be rather difficult to be performed once the peak is not `pure' though we try to limit the amplitudes of the given standard deviations $\{\sigma_{p}\}$.

%%%%%%%%%%%%%%%%%%%%%%%%%%%%%%%%%%%%%%%%%%%%%%%%%%%%%%%%%%%%%%%%%%%%%%%%%%%%%%%%
\begin{figure*}[htbp]
\centering
\includegraphics[width=0.75\textwidth]{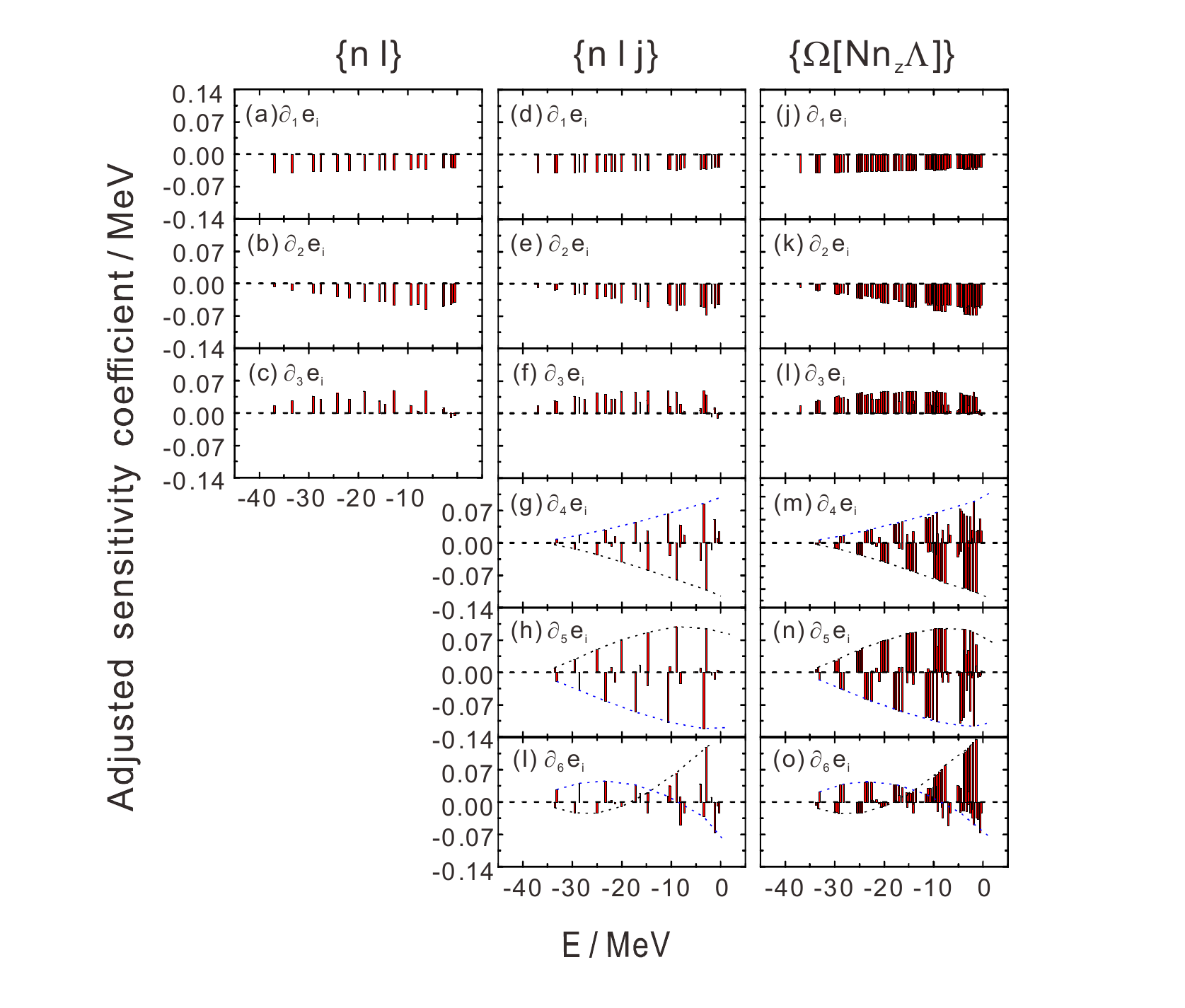}
\caption{
(Color online)
The adjusted sensitivity coefficients $\partial_je_i$ of three types of typical neutron single-particle levels labeled by $\{nl\}$,$\{nlj\}$ and $\{\Omega[Nn_z\Lambda]\}$ in $^{208}$Pb. See the text for more details.
                                                                   \label{fig7}
}
\end{figure*}
%%%%%%%%%%%%%%%%%%%%%%%%%%%%%%%%%%%%%%%%%%%%%%%%%%%%%%%%%%%%%%%%%%%%%%%%%%%%%%%%
%123456789 123456789 123456789 123456789 123456789 123456789 123456789 123456789
%description for Fig.7

Fortunately, we find that the single-particle energiy $e_i$ depends linearly on the potential parameters within the uncertainty domain near the universal parameters. That is to say, it is accuate enough to use the first-order Talor approximation in Eq.~(\ref{eqn.08}), which means that we can approximate the function $e_i=e_i(p_j)$ using its tangent line at the $p_j^{\rm o}$ point. Therefore, we can calculate analytically the energy uncertainty $\sigma_{e_i}$ according to Eq.~(\ref{eqn.08}). The partial derivatives (sensitivity coefficients) of the single-particle energies $e_i$ with respect to the potential parameters $\{p\}$ at $\{p^{\rm o}\}$ can be numerically calculated by the finite-difference formula,
\begin{equation}
   \frac{\partial e_i}{\partial p_j}
   \simeq
   \frac{e_i(p_j^+)-e_i(p_j^-)}{p_j^+-p_j^-},
                                                                   \label{eqn.16}
\end{equation}
with values of $p_j^+$ and $p_j^-$ suitably close to $p_j^0$. For convenience, we define an adjusted sensitivity coefficient as
\begin{equation}
   \partial_je_i
   \equiv
   \frac{\partial e_i}{\partial p_j}
   \sigma_{p_j}.
                                                                  \label{eqn.17}
\end{equation}
By giving a set of suitable $\{\sigma_p\}$, the adjusted sensitivity coefficients $\{\partial e_i\}\equiv\{\partial_1e_i,\partial_2e_i,\partial_3e_i,\partial_4e_i,\partial_5e_i,\partial_6e_i \}$  will have the similar order of magnitude.
Figure \ref{fig7} shows the adjusted sensitivity coefficients for the three tpyes of the calculated neutron single-particle levels labeled respectively by $\{nl\}$,$\{nlj\}$ and $\{\Omega[Nn_z\Lambda]\}$ in $^{208}$Pb. From this figure, it can be seen that the adjusted sensitivity coefficients show us the regular evolution trends. In particular, the spectrum envelopes, e.g., in Figs.~\ref{fig7}(g)-\ref{fig7}(i) and Figs.~\ref{fig7}(m)-\ref{fig7}(o), show different but interesting properties. It will be meaningful to reveal the physics behind them. Based on these sensitivity coefficients and the standard deviations of these model parameters or their combinations (namely, the adjusted sensivity coefficients), we can calculate the energy uncertainty $e_i$ analytically. Indeed, for the $i_{13/2}$ level, the analytical result coincides with the fitting value of the peak generated by the Monte-Carlo method. The typical error between the calculated and fitted values is less than 3\%.
%%%%%%%%%%%%%%%%%%%%%%%%%%%%%%%%%%%%%%%%%%%%%%%%%%%%%%%%%%%%%%%%%%%%%%%%%%%%%%%%
\begin{figure*}[htbp]
\centering
\includegraphics[width=0.6\textwidth]{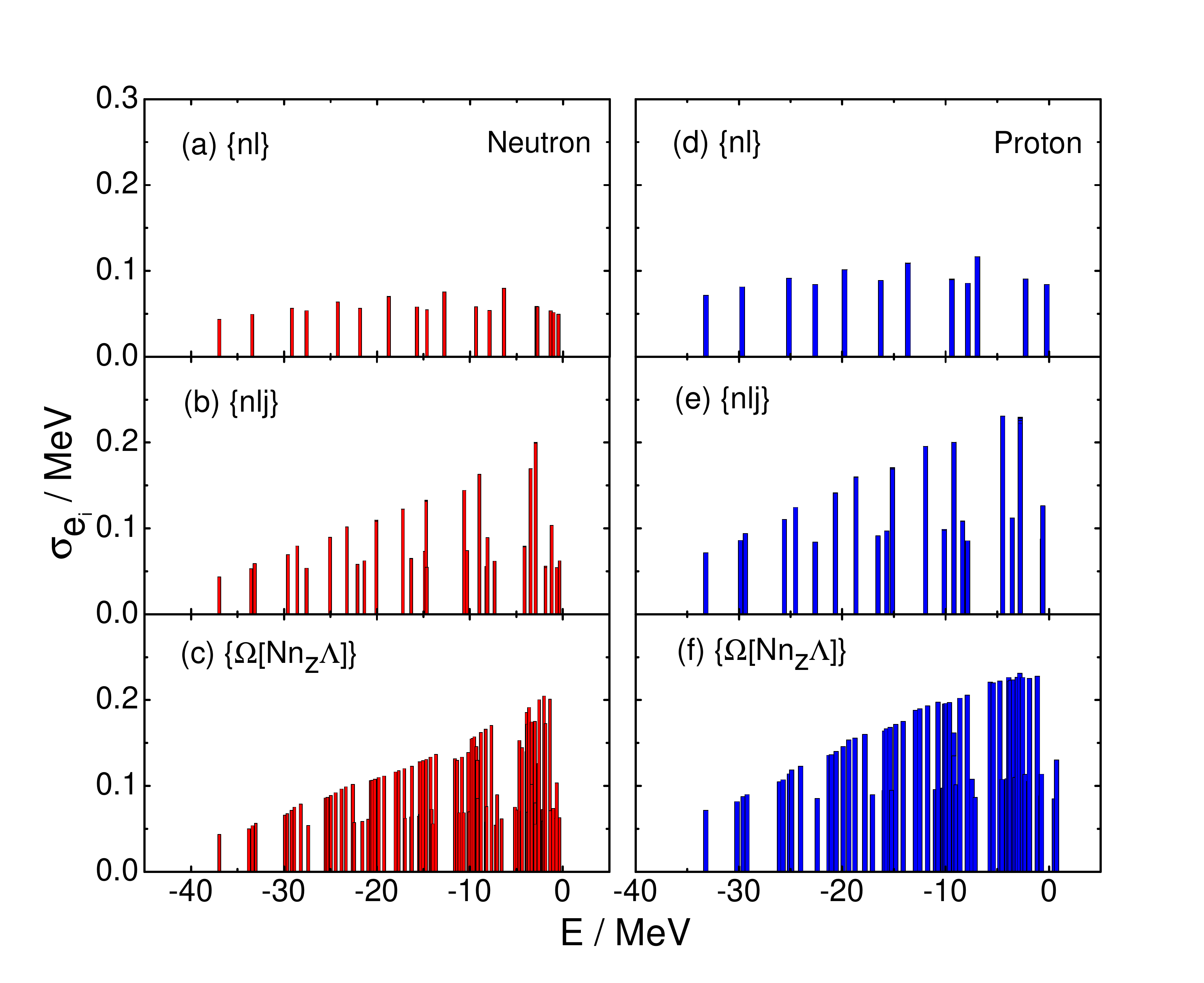}
\caption{
(Color online) The standard deviation $\sigma_{e_i}$ of three kinds of typically single neutron (left) and proton (right) energies in $^{208}$Pb.
                                                                  \label{fig8}
}
\end{figure*}
%%%%%%%%%%%%%%%%%%%%%%%%%%%%%%%%%%%%%%%%%%%%%%%%%%%%%%%%%%%%%%%%%%%%%%%%%%%%%%%%
%123456789 123456789 123456789 123456789 123456789 123456789 123456789 123456789
%description for Fig.8

Based on the above method, we analytically calculate the overall uncertainties of the three kinds of levels mentioned above both for neutrons and for protons in $^{208}$Pb. In the calculations, all the parameter uncertainties are taken into account, which indicates that calculations are done at  $\mathrm{s} =(1,1,1)$ point for  $\{nl\}$ levels and at $\mathrm{s} =(1,1,1,1,1,1)$ point for $\{nlj\}$ and $\{\Omega[Nn_z\Lambda]\}$ ones. As seen in Fig~\ref{fig8}, one can notice that the changing trends of the standard deviations are similar for neutrons and protons. For the $\{nl\}$ single-particle levels, there is no obvious change with changing $n,l$ quantum numbers or single-particle energies. For the $\{nlj\}$ and $\{\Omega[Nn_z\Lambda]\}$ single-particle levels, it seems that the increasing trends of the energy uncertainties appear with increasing energies or angular momentum $j$ for a given $n$, e.g, $n=1$. Note that one spherical $j$ mean-field orbital will split into $(j+\frac{1}{2})$ deformed substates e.g., at $\beta_2=0.1$ here. In addition, it was confirmed that one should obtain the same conclusions using the Monte Carlo technique. More attention in this work has been paid to the uncertainty propagation from the model parameter rather than more physical discussion behind it.

%%%%%%%%%%%%%%%%%%%%%%%%%%%%%%%%%%%%%%%%%%%%%%%%%%%%%%%%%%%%%%%%%%%%%%%%%%%%%%%%
\begin{figure*}[htbp]
\centering
\includegraphics[width=0.7\textwidth]{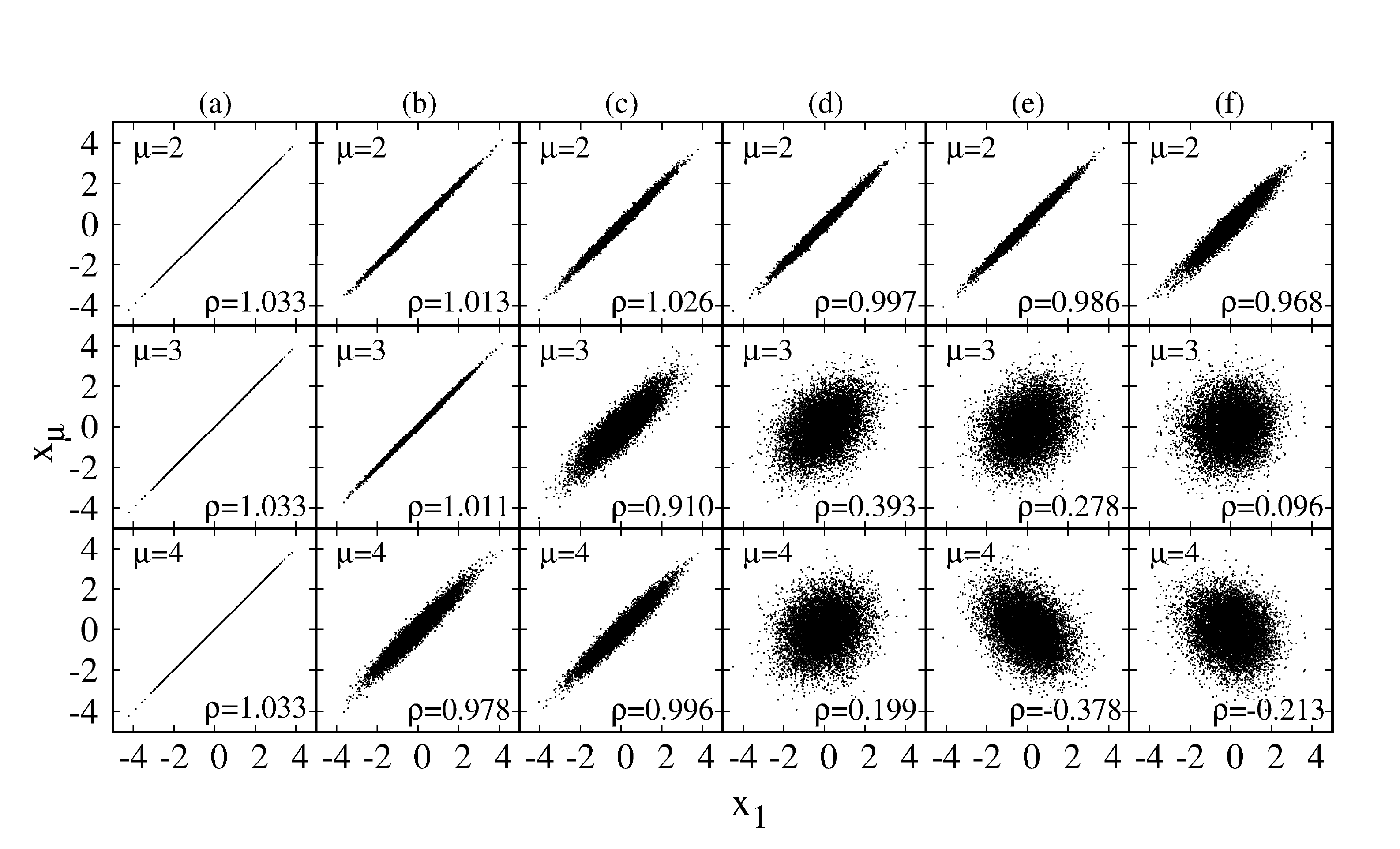}
\caption{
(Color online)  Scatter plots (with correlation coefficients $\rho$) between three pairs of arbitrarily selected single-neutron energies, $1i_{13/2}\oplus 1j{15/2}$ (top), $1i_{13/2}\oplus 3p{1/2}$ (middle) and $1g_{7/2}\oplus 1i{13/2}$ (bottom), near the Fermi surface in $^{208}$Pb. From the left to the right, the uncertainties of more and more parameters are considered as indicated in the text.
                                                                    \label{fig9}
}
\end{figure*}
%%%%%%%%%%%%%%%%%%%%%%%%%%%%%%%%%%%%%%%%%%%%%%%%%%%%%%%%%%%%%%%%%%%%%%%%%%%%%%%%
%123456789 123456789 123456789 123456789 123456789 123456789 123456789 123456789
%description for Fig.9

%%%%%%%%%%%%%%%%%%%%%%%%%%%%%%%%%%%%%%%%%%%%%%%%%%%%%%%%%%%%%%%%%%%%%%%%%%%%%%%%
\begin{figure*}[htbp]
\centering
\includegraphics[width=0.7\textwidth]{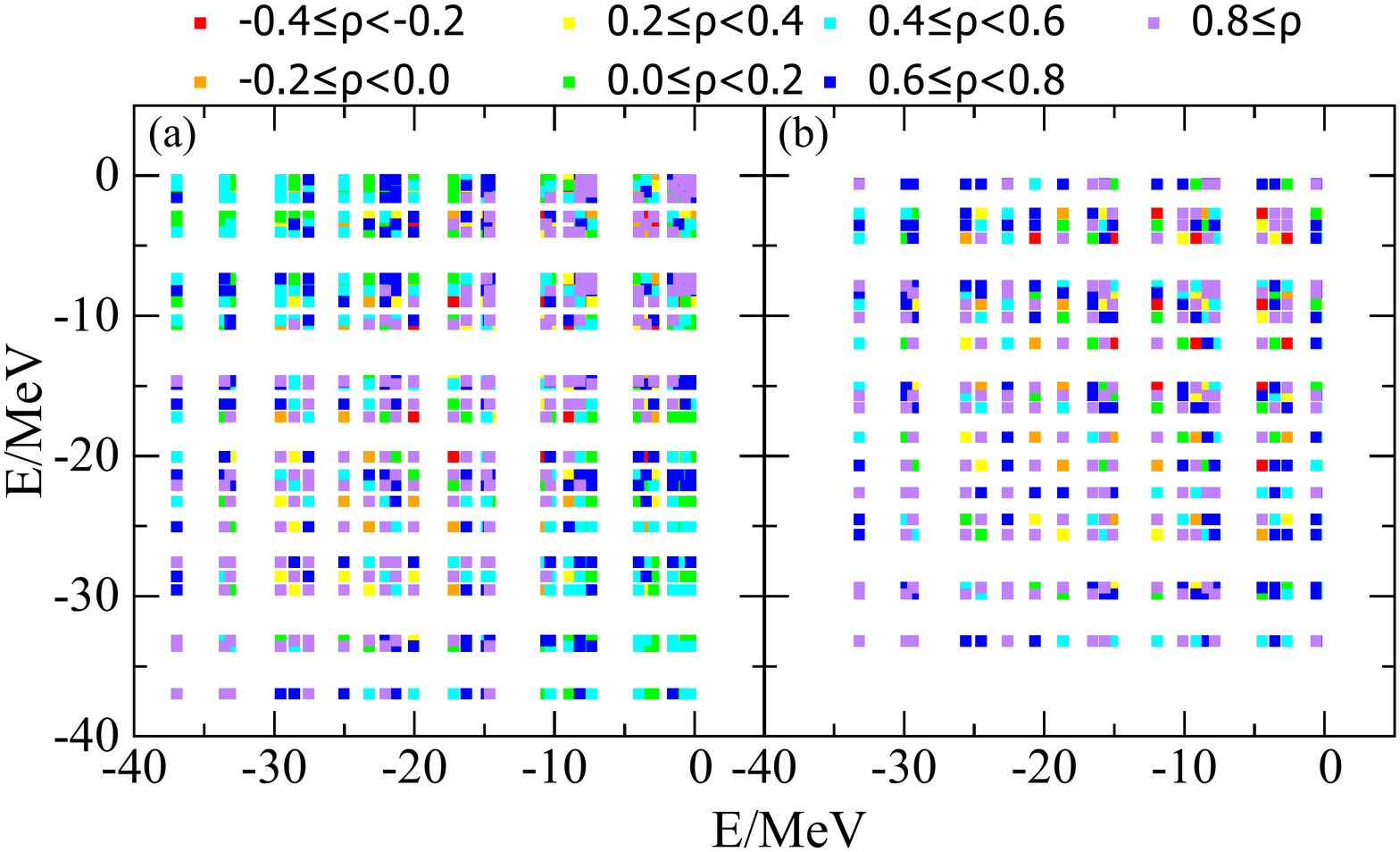}
\caption{
(Color online) Color-coded plot of the calculated correlation coefficients between single-particle energy levels for neutrons (a) and protons (b) in $^{208}$Pb.
                                                                   \label{fig10}
}
\end{figure*}
%%%%%%%%%%%%%%%%%%%%%%%%%%%%%%%%%%%%%%%%%%%%%%%%%%%%%%%%%%%%%%%%%%%%%%%%%%%%%%%%
%123456789 123456789 123456789 123456789 123456789 123456789 123456789 123456789
%description for Fig.9
%%%%%%%%%%%%%%%%%%%%%%%%%%%%%%%%%%%%%%%%%%%%%%%%%%%%%%%%%%%%%%%%%%%%%%%%%%%%%%%%
\subsection{Correlation Coefficients between single-particle energies}

As mentioned above, the single-particle levels are usually the input quantities in further theoretical calculations(cf., e.g., Refs.~\cite{Naderi2018,Chen2020}), e.g., the calculations of High-$K$ isomers, shell correction, pairing correction, etc. Both the energy uncertainties and the correlation effects are of importance for further uncertainty predictions. Based on Eq.~\ref{eqn.13} the Pearson correlation coefficients will be able to be calculated between any two levels. Further, we can investigate the correlation effects among them within the `small' energy domains associated with parameter uncertainties, $ \{\sigma_p\}$. Figure~\ref{fig9} shows the two-dimensional scatter plots between three pairs of arbitrarily selected $\{nlj\}$ single-neutron energies,$1i_{13/2}\oplus 1j_{15/2}$, $1i_{13/2}\oplus 3p_{1/2}$ and $1g_{7/2}\oplus 1i{13/2}$, near the Fermi surface. From the left to the right side in this figure, the calculations are performed at $\mathrm{s} =(1,0,0,0,0,0)$, $(1,1,0,0,0,0)$, $(1,1,1,0,0,0)$, $(1,1,1,1,0,0)$, $(1,1,1,1,1,0)$ and $(1,1,1,1,1,1)$ points, respectively. It should be noted that, similar to the operation in the Figure~\ref{fig2} plot, before plotting, the normal distributions of the selected single-particle levels are transfermed into the standard normal distributions by defining a dimensionless parameter,$x_i=[e_i({\mathrm{p}})-e_i({\mathrm{p}^{\rm o}})]/\sigma_{e_i}$. In Fig.~\ref{fig9}, the dimensionless parameters $x_\mu$, for $\mu=1, 2, 3$ and $4$, respectively correspond to the spherically mean-field single-particle levels $1i_{13/2}, 1j_{15/2}, 3p_{1/2}$ and $1g_{7/2}$. One can clearly see the evolutions of the correlation coefficients and the scatterplot distributions with opening more and more uncertainty parameters. In particular, it is found that the positive, zero, and negative values appear in the correlation coefficients.

%%%%%%%%%%%%%%%%%%%%%%%%%%%%%%%%%%%%%%%%%%%%%%%%%%%%%%%%%%%%%%%%%%%%%%%%%%%%%%%%
In order to give an overall investigation, we show that the color-coded plot of the calculated correlation coefficients between single-particle levels with energy $e_i < 0$ for protons and neutrons in Fig.~\ref{fig10}. As can be seen, the correlation coefficients have not the same values but cover a rather wide range. It is certainly important to consider these correlation effects when the single-particle levels with energy uncertainties are taken as input data to do the further calculations.

\section{Summary} \label{sec.IV}
Taking the $^{208}$Pb nucleus as the carrier, we have investigated the single-particle energy uncertainties and the statistical correlations of different levels due to the uncertainty propagation of independent model parameters, which are important for further theoretical predictions, e.g., the $K$ isomer calculation. The adjusted sensitivity coefficients are introduced and discussed for three types of single-particle levels. The overall standard deviations of the single-particle levels in the Woods-Saxon nuclear mean field are shown and the evolution properties are briefly discussed. It is also found that the correlation coefficients involve a rather wide domain, which are of importance for further theoretical uncertainty predictions relying on the single-particle levels. Noted that the practical energy uncertainties will depend on the practical standard deviations of model parameters during the further calculations whereas the evolution laws of parameter uncertainty propagations and the correlation properties of single-particle levels are still similar and valid. Next, we will further investigate the uncertainty propagation of model parameters with partial correlation effects. It would be also interesting to extend this study to other phenomenological or self-consistent models in nuclear physics, even other fields.
%%%%%%%%%%%%%%%%%%%%%%%%%%%%%%%%%%%%%%%%%%%%%%%%%%%%%%%%%%%%%%%%%%%%%%%%%%%%%%%%
%%%%%%%%%%%%%%%%%%%%%%%%%%%%%%%%%%%%%%%%%%%%%%%%%%%%%%%%%%%%%%%%%%%%%%%%%%%%%%%%
\acknowledgments
Some of the calculations were performed at the National Supercomputing Center in Zhengzhou. This work was supported by the National Natural Science
Foundation of China (Grant No. 11975209) and the Physics Research and Development Program of Zhengzhou University (Grant No. 32410017) and the Project of Youth Backbone Teachers of Colleges and Universities of Henan Province (Grant No. 2017GGJS008). 

%============================================================================================================================================
%============================================================================================================================================%

\appendices
\makeatletter % `@' now normal "letter"
\@addtoreset{equation}{section}
\makeatother  % `@' is restored as "non-letter"
\renewcommand\thesection{APPENDIX~\Alph{section}}
\renewcommand\theequation{\Alph{section}.\arabic{equation}}

%============================================================================================================================================%

%============================================================================================================================================%
\end{document}